arXiv:1208.0101

# Optimum inhomogeneity of local lattice distortions in La$_2$CuO$_{4+y}$

Nicola Poccia[1,2], Alessandro Ricci[1,3], Gaetano Campi[4], Michela Fratini[1,5], Alessandro Puri[6], Daniele Di Gioacchino[6], Augusto Marcelli[6], Michael Reynolds[2], Manfred Burghammer[2], Naurang Lal Saini[1], Gabriel Aeppli[7], Antonio Bianconi[1,8,9*]

[1]*Department of Physics, Sapienza University of Rome, P.le A. Moro 2, 00185 Roma, Italy*

[2]*European Synchrotron Radiation Facility, B.P. 220, F-38043 Grenoble Cedex, France.*

[3]*Deutsches Elektronen-Synchrotron DESY, Notkestraße 85, D-22607 Hamburg, Germany*

[4]*Institute of Crystallography, CNR, Via Salaria Km 29.300, Monterotondo Stazione, Roma, I-00015, Italy.*

[5]*Fermi Center, P.le del Viminale, 00187 Roma, Italy*

[6]*Istituto Nazionale di Fisica Nucleare, Laboratori Nazionali di Frascati, P.O. Box 13, 00044 Frascati, Italy.*

[7]*London Centre for Nanotechnology and Department of Physics and Astronomy, University College London, 17–19 Gordon Street, London WC1H 0AH, UK.*

[8]*Rome International Centre for Materials Science, Superstripes, Via dei Sabelli 119A. 00185 Roma Italy*

[9]*Mediterranean Institute of Fundamental Physics, Via Appia Nuova 31, 00040 Marino Italy*

*\*corresponding author*

*Classification:*

▪ *Physical Sciences, Applied Physical Sciences.*

**Abstract**:

Electronic functionalities in materials from silicon to transition metal oxides are to a large extent controlled by defects and their relative arrangement. Outstanding examples are the oxides of copper, where defect order is correlated with their high superconducting transition temperatures. The oxygen defect order can be highly inhomogeneous, even in "optimal" superconducting samples, which raises the question of the nature of the sample regions where the order does not exist but which nonetheless form the "glue" binding the ordered regions together. Here we use scanning X-ray microdiffraction (with beam 300 nm in diameter) to show that for $La_2CuO_{4+y}$, the "glue" regions contain incommensurate modulated local lattice distortions, whose spatial extent is most pronounced for the best superconducting samples. For an underdoped single crystal with mobile oxygen interstitials in the spacer $La_2O_{2+y}$ layers intercalated between the $CuO_2$ layers, the incommensurate modulated local lattice distortions form droplets anticorrelated with the ordered oxygen interstitials, and whose spatial extent is most pronounced for the best superconducting samples. In this simplest of high temperature superconductors, there are therefore not one, but two networks of ordered defects which can be tuned to achieve optimal superconductivity. For a given stoichiometry, the highest transition temperature is obtained when both the ordered oxygen and lattice defects form fractal patterns, as opposed to appearing in isolated spots. We speculate that the relationship between material complexity and superconducting transition temperature $T_c$ is actually underpinned by a fundamental relation between $T_c$ and the distribution of ordered defect networks supported by the materials.

---------------------------------------

**Introduction:**

Defects associated with lattice instabilities in solids are at the heart of many of their useful properties (1-3), including their electrical conductivity. For example, before the discovery of high-temperature superconductivity (HTS) in the cuprates, the search for new superconductors was influenced by the observation that the materials, such as $Nb_3Al$ ($T_c$ = 19 K), $Nb_3Ga$ ($T_c$ = 20 K), and $Nb_3Ge$ ($T_c$ = 23 K) (4), with the highest transition temperatures are at the edge of structural instability. Since then, lattice instabilities have been proposed to favor HTS (5). Indeed, the search for electronic-lattice instabilities of local structure in copper oxides was a driving idea for the discovery of HTS in the pseudo ternary oxide $La_{2-x}Ba_xCuO_4$ (6) and it was soon proposed that such materials were intrinsically phase separated (7). This was confirmed in $La_2CuO_{4+y}$, the simplest superconducting Cu oxide (with mobile oxygen interstitials) at the insulator to metal transition at very low doping (8). Further experiments revealed different phase segregation also in the optimum doping regime (9). The key role of lattice complexity follows because the critical temperature in all known HTS increases in compounds made of an increasing number of atomic elements as shown in Fig. 1. The lattice complexity of superconducting copper oxides was neglected by most popular theories of high $T_c$ superconductivity, while percolative theories for granular superconductors were invoked because of the relevance of lattice disorder for electronic properties (10).

Although surface defects, including their correlation with electronic properties, have long been investigated by surface-sensitive techniques such as scanning tunneling microscopy (STM) (11), bulk defect ordering has only recently become accessible. Here the important advance has been a novel experimental method taking advantage of synchrotron radiation focusing techniques: scanning x-ray micro-diffraction. It has been used for imaging phase separation in real space in cuprates, focusing on the self-organization of defects (12). A recent surprise has been that even for "optimal" samples in a single family, the ordering of oxygen interstitials is highly inhomogeneous (12). Nonetheless, while the best annealing protocol (13) in this instance did not yield homogeneous defect ordering, it did yield the most connected "fractal" network, thus suggesting that better superconductivity ($T_c = 40$ K) is due to percolation of regions with the best ordering of interstitials. Defect growth and annealing processes, in fact, determine the quality of the superconductivity through either reduction of their population or complex self-organization (14,15). The unanswered question concerns the nature of the medium hosting the regions with ordered interstitials. This is significant as it will play a key role in determining many generic properties, from mechanical and chemical stability to the electrical characteristics, both in the superconducting and normal states of many cuprates that do show the ordering of mobile interstitials and other defects during sample preparation. An example of such a property is the often observed linear relation between electrical resistivity and temperature, which could have an intrinsic, exotic origin in many-body physics but could also be specific to an inhomogeneous mixture.

For many years the dominant theories of HTS have considered a homogeneous stoichiometric $CuO_2$ layer and have neglected the key role of lattice effects (16,17), even if the $\left[ CuO_2 \right]_\infty$ layers are intercalated by a variety of defective oxide AO layers (A= La, Ba, Sr, Ca, Y, Hg or Rare earths) with a large tolerance factor (18) or misfit strain (19,20) and a large number of oxygen interstitials or defects. The lattice misfit induces tilting and corrugation of the $\left[ CuO_2 \right]_\infty$ layers and structural phase transitions going from I4/mmm (high temperature tetragonal, HTT), Bmab, Pccn, Fmmm (low temperature orthorhombic, LTO) to $P4_2$/ncm (low temperature tetragonal LTT) in systems like $La_2CuO_4$ (21).

Lattice effects, leading also to inhomogeneous internal strains, in perovskite manganites with fixed hole concentrations are well established and have been connected with their "colossal" magnetoresistance (22-24). It is therefore worth asking whether strain inhomogeneities are also an important feature of cuprates, where they could lead to apparent "phase segregation" (23,24) inducing peculiar transport properties, where the self-organization associated with long-range strain interactions could even induce electronic granular networks (25). Our recent discovery (12) that better superconductivity appears with "optimal inhomogeneity" (26) of oxygen defect ordering opens the question of the nature of the remaining disordered medium, the topic of the present paper. The separation into an ordered network and a strained embedding medium provides a basis for the two electronic components (27-29) which have been proposed as the key feature for certain theories of high temperature superconductivity (30-32).

The local lattice distortions of the $CuO_2$ planes (33-37) provide another kind of ordered defect. In several doped cuprates, they yield a characteristic incommensurate superstructure with the specific wave-vector 0.21**b**\* (38-46). Recent STM work has clearly shown that the local lattice distortions of the $CuO_2$ plane induces a spatial modulation of the gap (11). The incommensurately modulated local lattice distortions (LLD), whose density increases below 200 K appear in droplets of several nanometers across, as indicated by the widths of diffuse x-ray satellites (37-45). Here we use X-ray microscopy to show that for the cuprate superconductor, La$_2$CuO$_{4+y}$, the dominant features of the embedding "glue" regions, where interstitials are disordered, are optimally ordered arrays of tilt defects of the copper-centred oxygen plaquets, which are the fundamental building blocks of the cuprates.

## Materials Aspects

La$_2$CuO$_{4+y}$ is the simplest cuprate and therefore represents an ideal venue for investigation of LLD droplets. For $0.01 < y < 0.055$, La$_2$CuO$_{4+y}$ shows macroscopic phase separation between the y1 = 0.01 antiferromagnetic phase and the y2 = 0.055 superconducting phase (8,41), where the sample shows the coexistence of two c-axis lattice constants corresponding to two competing macroscopic phases. For y>0.055, La$_2$CuO$_{4+y}$ shows more subtle phase separation, characterized by coexistence of two different superconducting phases with different critical temperatures, the first $T_{c1} = 15 \pm 1 \ K$, and the second T$_c$ ranging between 27 and 38 K, even while the crystalline lattice shows no splitting of the c-axis. The first critical temperature is sensitive to O$i$ ordering, and the second critical temperature is similar to that seen for La$_{2-x}$Sr$_x$CuO$_4$ (11, 46,47).

We have first investigated the competition among O$i$ puddles and LLD droplets, (see table 1 for the acronyms) using a La$_2$CuO$_{4+y}$ single crystal in the underdoped regime with y=0.06, which corresponds to an electronic doping of 0.1 holes per Cu site. The superconducting critical temperature and the pinning properties have been characterized by ac-susceptibility experiments. The sample shows electronic phase separation into two superconducting phases with two critical temperatures Tc = 14 K and T$_c$ = 27 K in agreement with previous studies (12,47,48). Synchrotron X-ray diffraction (XRD), performed at the European Synchrotron Radiation Facility (ESRF) at Grenoble, shows weak incommensurate diffuse satellites (called Q3-LLD) displaced by incommensurate wave-vector **q3** = 0.21 **b**\*+0.29 **c**\* from the principal Fmmm lattice reflections (see figure S3) for this mono-crystal. The diffuse weak Q3-LLD satellites coexist with Q2-O$i$ satellites (displaced by **q2** = 0.25 **b**\*+0.5 **c**\*) due to O$i$ sitting at the (¼,¼, ¼) interstitial site position (49). Q3-LLD satellites dominate and are easier to detect in underdoped samples. On the contrary, Q2-O$i$ satellites are dominant at the optimum doping, for y > 0.1.

**Results and Discussion:**

 For our scanning X-ray microdiffraction experiments, we used the ID13 beam line of ESRF, optimized to deliver X-rays with an energy of 14 keV, focused on a 300 nm spot on the sample surface. Data were collected in the reflection geometry with a two-dimensional Fast Readout Low Noise Charged Coupled Device (FReLoN CCD) detector. We constructed images from 6370 XRD diffraction patterns, each one for a different spatial x-y position of the sample.

Figures 2A and 2B contain schematics of the two coexisting structural modulations in the bc plane, the Q2-Oi and Q3-LLD, respectively. Figure 2C is a three-dimensional image of O$i$ puddles and LLD droplets. We may recognize the isolated puddles (cold color) and the dominant droplets phase (hot color). Fig. 2D demonstrates that the two signals are anticorrelated. The results show that the ordered oxygen interstitials give a granular superconducting phase with $T_c$ = 14 K in the underdoped regime competing with the second superconducting phase made of a distribution of droplets of LLD with $T_c$ = 27 K. Figure 2E shows a pictorial scheme of the spatial distribution of the LLD droplets (blue circles) with a 23 nanometer size, deduced from the diffraction line-width (see supplementary materials Fig. S2), and of the large O$i$ puddles (red polygons). The imaging of mesoscopic spatial inhomogeneity points clearly toward the assignment of the superconducting phases in La$_2$CuO$_{4+y}$ with $T_c$ = 14 K in this sample and with $T_c$ = 40 K in the optimum doped sample (10) to the ordering of mobile O$i$ forming isolated puddles and their scale-free pattern organization respectively.

We have been able to obtain a sample showing only Q3-LLD satellites by disordering the O$i$ via heat treatments (11,12) as shown in Figure 3. We increased the sample temperature above the O$i$ order-to-disorder temperature, i.e., 350 K, followed by a rapid quench below 200 K. In such a way, O$i$ remains frozen in a disordered state and the Q2-O$i$ diffraction peaks, as measured by the CCD area detector, are completely missing (Fig. 3A). Fig. 3B is the scanning X-ray microdiffraction image of the pattern of LLD droplets in this sample with $T_c$ = 27 K.

To identify how LLD droplets can arise we carried out experiments of photo-induced effects at the Trieste synchrotron radiation facility Elettra (13). The X-ray beam spot size was 200 μm$^2$ on the sample surface, and the flux (defined as the number of photons hitting the sample surface per second per unit area) $\Phi_{P(0.1nm)} = 5 \times 10^{14} N_{P(0.1nm)} s^{-1} cm^{-2}$ corresponds to a power density of 1 $W\ cm^{-2}$ on the sample surface. The X-ray photon beam is at the same time a pump and probe excitation of a surface layer of about 1.5 μm. The effect of continuous illumination corresponds to photo-excitation in which the state-changing rate is proportional to the intensity of the radiation. Therefore, the physical state of the system is controlled by the fluence $F_{P(0.1nm)} \left( N_{P(0.1nm)} \cdot cm^{-2} \right) = \Phi_P \cdot t$. The sample was kept at constant temperature T = 85 K where the O$i$ are frozen (12,13) in the disordered glassy phase of the quenched sample. Figure 3C shows the time evolution of the intensity of the Q3-LLD satellites recorded in the CCD area detector probing the (b*c*) reciprocal space at the fixed temperature of 85 K. The time evolution of the Q3-LLD XRD intensity follows the equation $I(t) \propto \left( 1 - e^{-\frac{t}{\tau}} \right) \cdot t^\gamma$. During experiments we checked that upon doubling or halving the flux, the timescales were halved or doubled, respectively; therefore the variation of the Q3-LLD intensity is plotted as a function of the fluence. The power law regime for the droplets follows the increase of the XRD diffraction intensity, without threshold, with an exponent $\gamma = 0.1 \pm 0.02$. To determine the temperature range where the X-ray illumination stimulates the Q3-LLD ordering, we have performed a thermal cycle from 100 K to 400 K under a high flux illumination. The Q3-LLD satellite intensity as a function of the temperature obtained by heating the sample after the X-ray illumination at 85 K is shown in Fig. 3D. The number of ordered domains Q3-LLD, proportional to the integrated intensity of the reflections Q3-LLD, returns to its initial value after increasing the temperature above 200 K. The results are in agreement with i) photo-annealing of the incommensurate modulated local lattice distortions detected by electron diffraction peaks below 200 K (43); ii) the photo-induced variation of superconducting properties (50,51); and iii) with the onset of LLD below 200 K as detected by EXAFS (43).

The LLD droplets form networks whose nature varies with superconducting critical temperature. We have used X-ray micro-diffraction apparatus at the ESRF to map the evolution of the Q3-LLD satellites for five single crystals of electrochemically doped La$_2$CuO$_{4+y}$, from the underdoped state (y = 0.06) to the optimum doping range, 0.1<$y$<0.12. Figure 4A and 4B show respectively the probability distribution of XRD Q3-LLD intensities and the spatial correlation function, $G(r)$, where $r = |\mathbf{R}_i - \mathbf{R}_j|$, calculated for the intensities at the spots $\mathbf{R}_k$. From the XRD mapping we have extracted the probability distribution, P(x), of the intensity I(Q3) of the reflections due to XRD Q3-LLD satellites of the main crystalline reflections, and normalized to the background, I(Q3)/I$_0$. Normalized data have been divided by the mean XRD intensity of the sample, x, and scaled using a power law with a cut-off x$_0$: $P(x) \propto x^{-\alpha} \exp(-x/x_0)$ with the power-law exponent $\alpha = 2.6 \pm 0.1$. All probability distributions of XRD intensities scale with the same power-law exponent $\alpha$ and a variable cut-off in the range from 4.5 to 15 (Fig. 4A). The spatial correlation function follows a power law, $G(r) \propto r^{-\eta} \exp(-r/\xi)$, with the exponent $\eta = 0.3 \pm 0.1$ and the correlation length $\xi = 30 \pm 10$ $\mu m$ in the underdoped y=0.06 sample with T$_c$ = 27 K.

Looking at the distance-dependent intensity correlations from the underdoped to the optimum doping state indicates that the droplets self-organize in a fractal state. Fractals appear in many fields [52-54], including all branches of materials science where new phases grow via stochastic nucleation and accretion at first order phase transitions [55]. Of course the detailed nature, including characteristic exponents will depend strongly on the details, such as annealing protocols and strain interactions, for the particular system under consideration. For the LLD regions in La$_2$CuO$_{4+y}$, the correlation length follows a power law with an exponential cut-off which progressively grows with higher critical temperatures. Figure 4B shows that G(r) for samples with T$_c$ in the range 27 K >$T_c$> 38 K, collapse onto the same curve when plotted versus r/$\xi$. In particular, for 30 < T$_c$ < 35 K, $\xi$ is in the range of 40 to 120 $\mu m$ and for T$_c$ = 37 ± 1 K it is 140 ± 20 $\mu m$. At the same doping level, therefore, there are shorter correlation lengths for the LLD droplets than the O$i$ puddles, which in the optimal case can reach even 400 $\mu$m (12,13).

Figure 5 shows the T$_c$ in the range 27-38 K associated with the droplet network, as a function of the cut-off of the probability distribution of XRD Q3-LLD intensities. The critical temperature scales with the cut-off according to a power law with an exponent 0.4 ± 0.05. This result points again toward the importance of connectivity and an optimum inhomogeneity for high critical temperature (12,26). It is also in qualitative agreement with the theoretical prediction of the increase of T$_c$ in a granular superconductor on an annealed complex network with a finite cut-off (56,57). In fact, for a power law distribution of links in a granular superconductor with an exponent $\alpha = 2.6$ the critical temperature is predicted to increase as a function of the cut-off with an exponent $3 - \alpha$, as observed experimentally. Therefore these results support the new physics of quantum phase transitions for fermions (56,57) and bosons (58) on scale free networks.

**Conclusions:**

We demonstrate that $La_2CuO_{4+y}$ actually contains networks of two superconductors characterized by different ordered defects (O$i$ and LLD). The best fractal behavior and superconductivity is obtained simultaneously for both O$i$ and LLD order. In particular, we have provided a positive correlation between the cut-off, $x_o$, for scale-invariance of the LLD diffraction intensity distribution and $T_c$. Furthermore, the strains in the LLD droplets are correlated over the longest distances when the stresses produced over still larger distances by the ordered interstitials display their maximal correlations, a condition which also yields superconductivity with a maximum $T_c$. Therefore we argue that the best fractal O$i$ network strains the embedding medium with LLD order.

Our quantitative results should be tested against theories of composite, granular superconductors proposed for cuprates (10,14,15,25,30-32,59-62). The X-ray data indicate that these theories must take into account not only the usual superconducting proximity effects, but also the effects of strains which the two components exert on each other. It is the latter which must be ultimately responsible for the exponents $\alpha$ and $\eta$ which we observe, and given the long-range nature of strain interactions, they cannot be accounted for within a simple near-neighbor percolation model. Our work provides a rationale for the observation of Fig. 1 showing that the main determinant of $T_c$ is complexity of the underlying material. Indeed, having already shown that there are at least two highly relevant, co-existing networks of ordered defects in the simplest cuprate, it is quite possible that there are numbers N>2 of such networks for the more complex materials. $T_c$ then grows with N because of two effects: (i) the larger chances of optimal strain and Josephson (proximity) couplings with increasing N, even though the underlying nanoscale pairing propensities remain invariant, and (ii) the accommodation of larger doping densities in portions of samples with larger N. Our thinking suggests a clear program for future theory, taking into account random network, long range strain and granular superconductivity concepts, and experiments, exploiting X-ray microdiffraction to identify the order parameters and microstructure of these networks not only for the cuprates, but also for other complex superconductors, such as the pnictides for which analogous phase separation effects have been recently observed (63-65).

**Author notes:** N.P., A.R., M.F. G.C., A.B., N.L.S. have performed the experiments and followed the data analysis; A.P., N.P., D.dG., A.M., have done the transport measurements; M.R., M. B., have provided the XRD station at ESRF; A.B., A.M., N.L.S., and N.P. have planned the experiment, the data analysis and together with G.A. have written the paper. These experiments were part of a program originated by A.B. and G.A. under the COMEPHS EU FP6 project.

**Supplementary Information is available at:**

66. Table 1

| acronyms | |
|---|---|
| O$i$ | Oxygen interstitials sitting at the sitting at the (¼,¼, ¼) interstitial site position in the $K_2NiF_4$ lattice structure |
| Q2-Oi | Self organization of oxygen interstitials with superlattice wave-vector $\mathbf{q2} = 0.25\ \mathbf{b}*+0.5\ \mathbf{c*}$ |
| LLD | Local Lattice distortions |
| Q3-LLD | Self organization of local lattice distortions with superlattice wave-vector $\mathbf{q3} = 0.21\ \mathbf{b}*+0.29\ \mathbf{c*}$ |

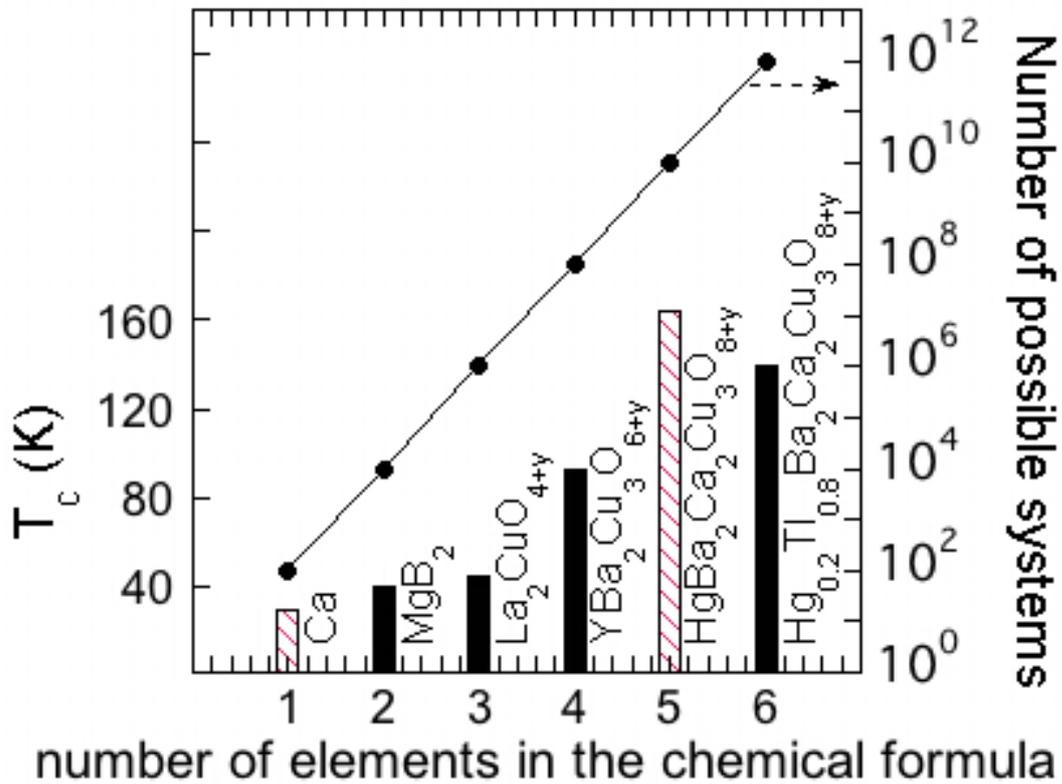

**Fig.1**: The maximum superconducting critical temperature, discovered so far by materials research, increases by increasing the lattice complexity. The maximum critical temperature (at ambient pressure (solid bars) and high pressure (dashed bars) in systems made by a single element (Ca under pressure) and two elements (magnesium diboride), and copper oxides with multiple elements in the chemical formula.

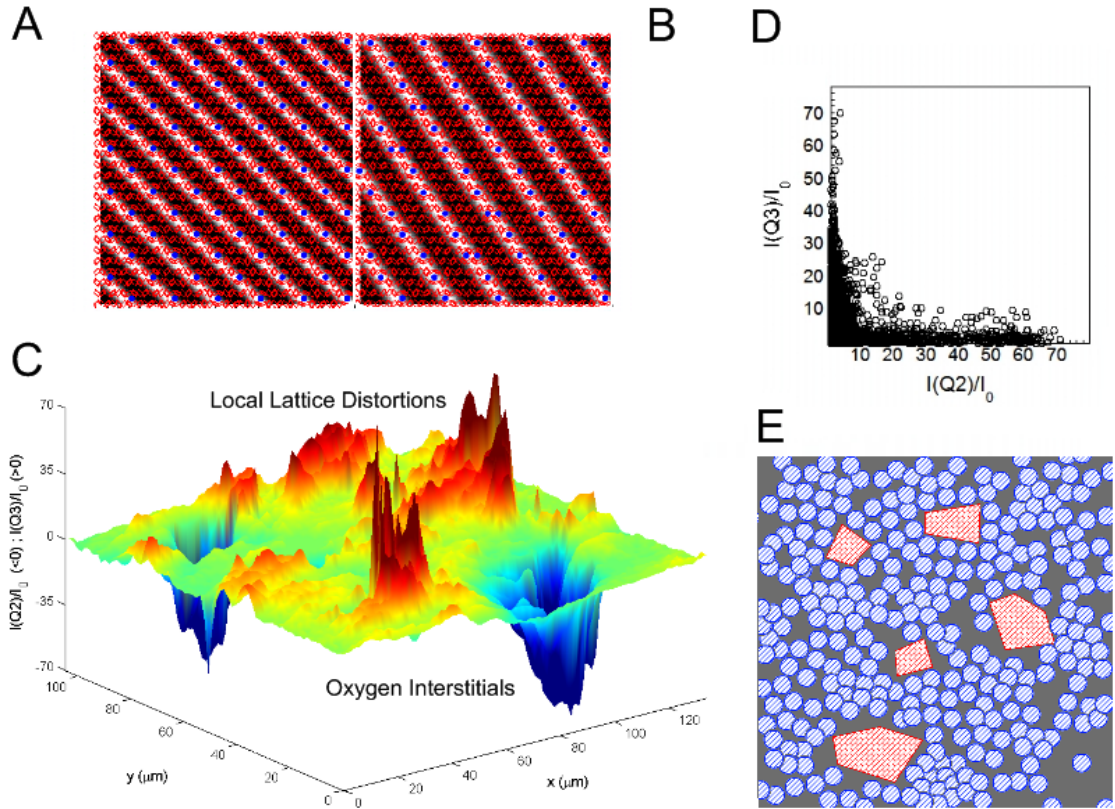

**Fig. 2**. The coexistence of the local lattice distortions (LLD) droplets and ordered oxygen interstitials (Oi) puddles in different spatial locations of $La_2CuO_{4+y}$ for y ≈ 0.06 as seen by scanning X-ray microdiffraction. The pictorial view of the Q2-Oi puddles made of Oi ordered with Q2 superstructure (panel **A**) and of the Q3-LLD puddles made of ordered LLD with Q3 superstructure (panel **B**) in the **bc** crystal plane of the Fmmm structure of $La_2CuO_{4+y}$ **C.** The three dimensional color plot imaging the position dependence of the Q3-LLD superstructure intensity $I(Q3)/I_0$ (values > 0) and of the Q2-Oi superstructure intensity $I(Q2)/I_0$ (values < 0). The scanning images show a few large disconnected Q2-Oi islands (negative blue-dark valleys) embedded in a matrix of the granular superconductor made of Q3-LLD (positive red-dark peaks). Data have been then normalized to the intensity ($I_0$) of the tail of the main crystalline reflections at each point (x, y). Visual inspection of both the mapping x–y position dependence of the integrated satellite peak intensity for Q2-Oi and Q3-LLD shows that from the scale of hundreds of nanometers to micrometers, the ordered Oi and the ordered LLD occupy distinct locations in space. The intensities of the superstructure satellites due to Q3-LLD and Q2-Oi ordering have been integrated over square sub-areas of the images recorded by the CCD detector in reciprocal-lattice units (r.l.u.). **D.** The Q3-LLD superstructure intensity $I(Q3)/I_0$ and of the Q2-Oi superstructure intensity $I(Q2)/I_0$ are plotted as a function of each other. The resulting plot indicates a high degree of anti-correlation between the two type of domains characterized by different superstructures. **E.** The schematic view of the spatial distribution of the LLD droplets (blue circles) and the ordered Oi puddles (red polygons). The grey backgrounds are regions of the sample where neither droplets or puddles are present.

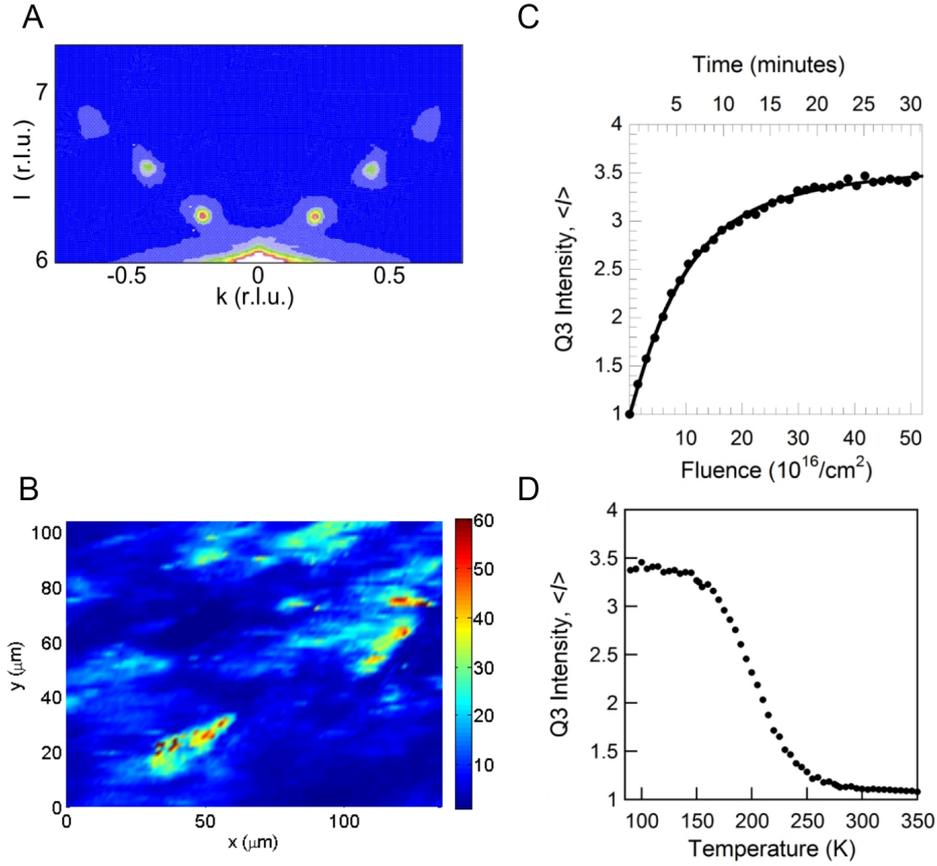

**Fig. 3. A.** The CCD image of the Q3-LLD satellite in the b*-c* plane near the main Fmmm reflections of the underdoped La$_2$CuO$_{4+y}$ single crystal, after the removal of the Q2-O$i$ satellite by rapid quenching after heating the sample above 350 K. Crystals are cooled to liquid nitrogen temperatures (as low as 85 K) with a 700 series Oxford Cryosystems cryocooler. **B.** The position dependence of the Q3-LLD superstructure intensity $I$(Q3)/$I_0$ in the two dimensional colour plots after the removal of the Q2-O$i$ superstructure intensity $I$(Q2)/$I_0$ by thermal annealing. **C.** The intensity of the Q3-LLD XRD reflections is plotted as function of fluence $\phi$ or time for constant X-ray flux. The surface is illuminated by a X-ray flux $\phi_{p(0.1\ nm)} = 5 \cdot 10^{14}\ N_p \cdot s^{-1} cm^{-2}$ keeping the temperature constant at 85 K. **D.** The temperature evolution of the Q3-LLD satellite intensity in the range 85 - 350 K collecting images every 2 K. The time evolution experiment has been carried out at the Elettra storage ring in Trieste. The X-ray beam, emitted by the wiggler source was monochromatized at the 0.1 nm wavelength by a Si(111) double crystal monochromator and focused on the sample surface at the X-ray diffraction beamline (XRD1).

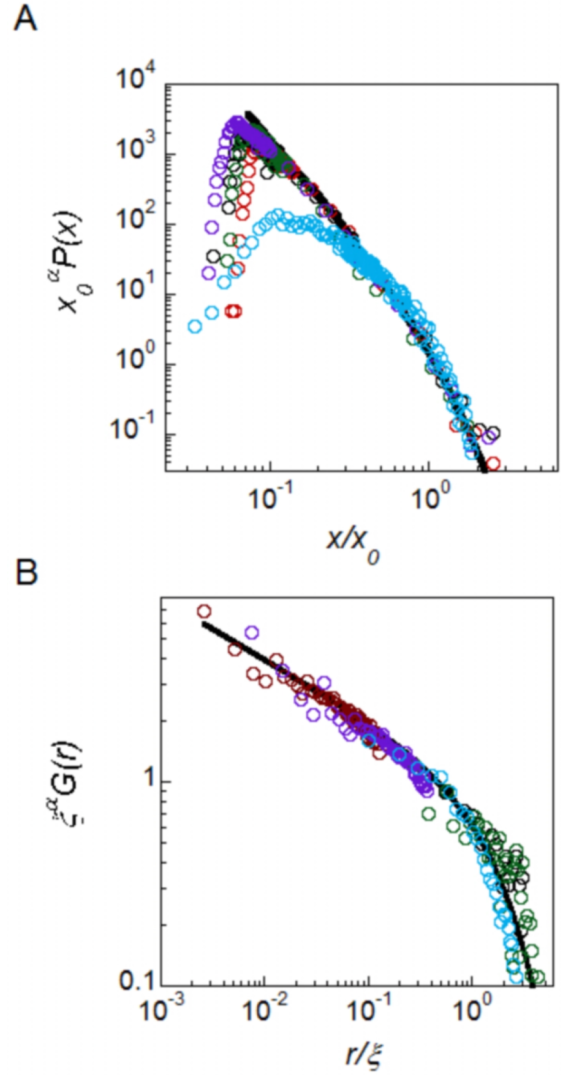

**Fig. 4. A.** The probability distributions, P(x), of the Q3-LLD XRD intensity x = I(Q3)/I$_0$ for single crystals of electrochemically doped La$_2$CuO$_{4+y}$ from the underdoped state (y = 0.06) to the optimum doping range, $0.1 < y < 0.12$. The curves follow a power law distribution $P(x) \propto x^{-\alpha} \exp(-x/x_0)$ with a variable exponential cut-off x$_0$. The curves $x_0^{\alpha} P(x)$ of all samples as a function of x/x$_0$ collapse on the same curve. **B.** The spatial correlation function, G(r), of the Q3-LLD XRD follows a power law distribution $G(r) \propto r^{-\eta} \exp(-r/\xi)$. The correlation length ξ varies from 30 to 140 m increasing with the doping range of the material investigated. The curves $\xi^{\alpha} G(r)$ of all samples as a function of $r/\xi$ collapse onto the same curve.

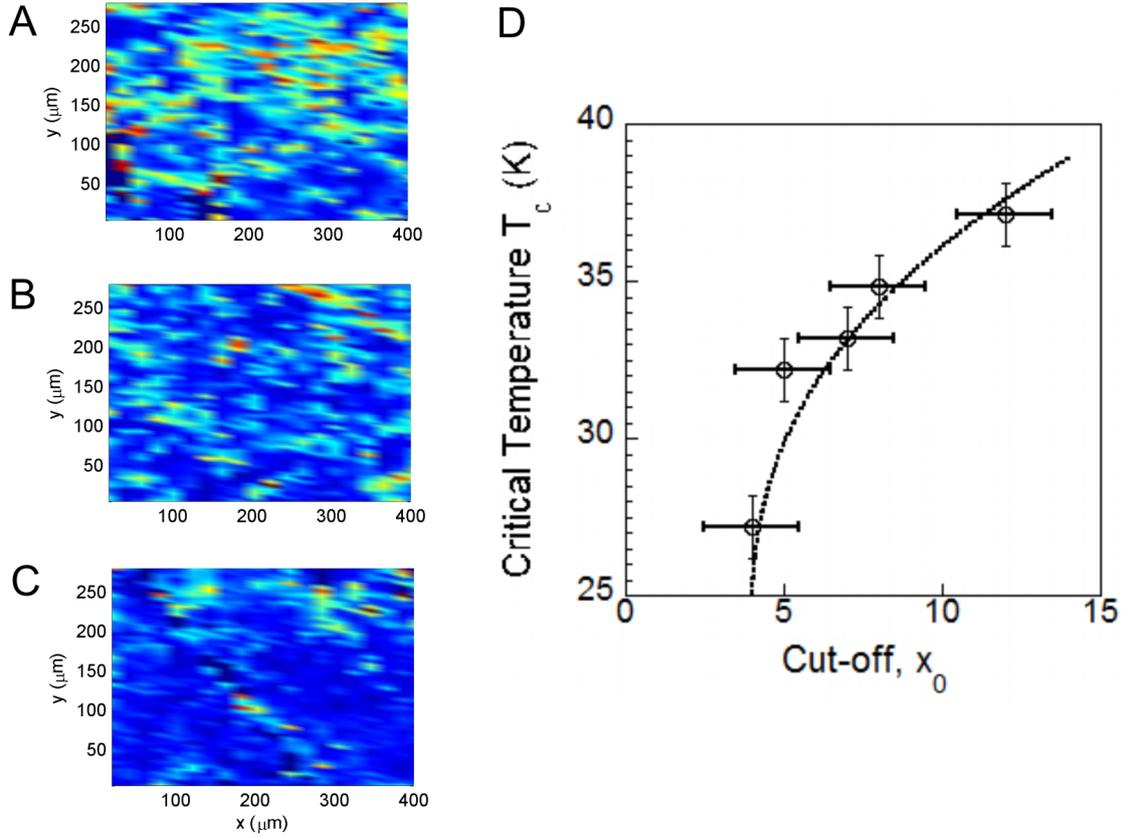

**Fig. 5. A,B,C.** X-ray microdiffraction results for the position dependence of the Q3-LLD superstructure intensity $I$(Q3)$/I_0$ in the La$_2$CuO$_{4+y}$ crystals with different critical temperature, T$_c$, 37, 34 and 32 K from A to C. The scanning XRD images show the better self organization of LLD droplets, proceeding to the higher T$_c$. **D.** The critical temperature T$_c$ in the range 25 K < T$_c$ < 37 K for five samples is plotted as a function of the cut-off parameter of the distribution of the LLD droplets density probed by the intensity distribution of the Q3-LLD superstructure satellites. Error bars in the critical temperature are of ± 1 K. The dashed line is the fit with a power law curve with exponent 0.4 ± 0.05, in agreement with theoretical predictions in reference [56] for granular superconductivity on a scale invariant network.